\documentclass[preprint,rmp,aps]{revtex4}
\usepackage{graphicx}
\usepackage{dcolumn}
\usepackage{bm}

\begin{document}

%



\title{A Density Functional Study of Magnetism in Bare Gold Nano-clusters}

\author{R.   J.   Magyar}
\affiliation{INEST\footnote{Interdisciplinary Network of Emerging Science and Technologies} 
Group Postgraduate Program, \\
Philip Morris USA ,  4201 Commerce Road,
Richmond, VA 23234
\\and\\
NIST Center for Theoretical and Computational Nanosciences (NCTCN),  
100 Bureau Dr. MS 8380, Gaithersburg, MD 20899}

\author{V. Mujica}
\affiliation{INEST Group Postgraduate Program,
Philip Morris USA ,  4201 Commerce Road,
Richmond, VA 23234, \\
Argonne National Laboratory, Center for Nanoscale Materials, Argonne, IL 60439 
\\and\\
Northwestern University
Department of Chemistry
2145 Sheridan Road
Evanston, IL 60208
}

\author{M. Marquez}
\affiliation{Research Center, Philip Morris USA, 4201 Commerce Road,
Richmond, VA 23234, \\
Harrington Department of Bioengineering, Arizona State University,  
Tempe, AZ 85280
\\and\\
NIST Center for Theoretical and Computational Nanosciences (NCTCN),  
100 Bureau Dr. MS 8380, Gaithersburg, MD 20899}

\author{C. Gonzalez}
\affiliation{
Computational Chemistry Group and
NIST Center for Theoretical and Computational Nanosciences (NCTCN),  
100 Bureau Dr. MS 8380, Gaithersburg, MD 20899}

\preprint{To be submitted to PRB}


\begin{abstract}

Magnetism in bare uncapped gold nano-clusters is explored from a density functional theory perspective with scalar
relativistic effects included via the pseudo-potential. The computed electronic structures of various nano-clusters reveal that permanent size-dependent spin-polarization appears without geometry relaxation for bare clusters even though bulk gold is diamagnetic.  The polarized ground states for 
clusters are favorable due to the
hybridization of the s and d orbitals, and bare octahedral clusters are expected to be magnetic for cluster sizes of approximately 38
atoms and larger. Much larger clusters will be diamagnetic when the surface-to-volume ratio is small and the core
diamagnetism prevails.   Moderate changes in the inter-atomic distances and cluster geometry are shown not to alter this conclusion.   Contrary to LDA and EAM predictions, GGA and hybrid geometry optimizations 
reveal increased inter-atomic bond distances in bare gold clusters relative to the bulk lattice values.   This expansion enhances the preexisting spin polarization.  

\end{abstract}


\maketitle


\section{Introduction}
\label{s:intro}

Modern synthesis and characterization techniques have enabled the controlled production of nano-sized metal and
semi-conductor clusters.  Because of their small size, these nano-clusters exhibit adjustable optical and physical
properties that may differ drastically from their bulk properties.  Since these properties often depend critically upon
the size and shape of the nano-cluster, it is possible that through synthesis, nano-clusters can be tuned to be suitable
for precise technological applications. In particular, it would be highly advantageous to fashion nano-particles with 
tunable optical, magnetic, and electron transport properties.  This combination would be particularly useful in targeted drug delivery schemes as well as for imaging and diagnosis biomedical applications. 
For example in one scheme, the nano-particles could be injected into a patient, and then, the concentration of the
nano-particles could then be localized in a region of interest by an external magnetic field. Finally, radiation tuned to
the particles' plasmon modes could thermally activate the particles resulting in local heating that could for example kill
undesirable cells \cite{S4}.   Gold forms a nano-particle with promising optical and magnetic properties \cite{DA4}.  It is
enticing for medical applications because it is both easy to manipulate and is generally considered to be chemically inert. 

Despite conflicting experimental results, there is a consensus that gold nano-particles of with approximately hundreds of atoms can retain a
magnetic moment under zero field bias.  Furthermore, their plasmon range is proportional to the size of the nano-particle \cite{DA4}. 
If both of these conditions are satisfied and controllable, gold would be an even stronger candidate for the afore
mentioned applications.  Understanding what causes this magnetism will be required to exploit, with full advantage, this magnetic behavior in future applications. 

Gold magnetism was first observed by Zhang
and Sham who used X-ray spectroscopy techniques on alkane-thiolated nano-clusters \cite{zs3}.  A subsequent experiment by Crespo and coworkers used X-ray absorption near-edge structure (XANES) to
show that, in thiolated clusters with on the order of a hundred atoms, the gold atoms exhibit a magnetic moment per atom of $\mu=0.036 \mu_B$
\cite{clr4} where $\mu_{B}$ is the Bohr magneton (9.27$\times 10^{-24}$ J / T).  In both experiments, the magnetism was
detected when sulfur-based capping ligands were used, and this lead Crespo and coworkers to conclude that ligand-cluster
interactions might play a vital role in inducing magnetism.  Based on this hypothesis, optical properties could be
controlled by choosing the size of the cluster while magnetic properties seemed to be the result of the coverage of sulfur-based capping
ligands.  It should be noted that neither of these experiments were able to observe magnetism in gold clusters capped with weakly-interacting nitrogen-based ligands instead of strongly-interacting sulfur-based ligands such as thiolates.  This negative result  is seemingly in contradiction with the results reported by Yamamoto and coworkers obtained using X-Ray
Magnetic Circular Dichroism spectroscopy (XMCD) \cite{yms4}.  They found that the presence of magnetism depended strongly
on the type of the surfactant and that the magnetism was much stronger for nitrogen based ligands.  They claimed that
the strong covalent bond between sulphur and gold induces a spin singlet state and actually quenches magnetism.  They concluded
that gold clusters are intrinsically magnetic and that capping ligands could quench this.  Yamamoto and
coworkers went on to measure the magnetic moment as a function of cluster size showing that magnetism is indeed a
size-tunable property in these systems \cite{hyi4}. 

There have already been several attempts to model gold nano-clusters theoretically from first principles.  The difficulty
with these theoretical techniques is many fold.  First of all, it is unclear how well these nano-clusters are
experimentally characterized in terms of homogeneity, size, and consistency.  Second, given the size of a cluster, the
exact arrangement of the atoms is uncertain.  Third, the achievable accuracy of the calculational methods is constrained
by the large number of atoms that must be considered to approach typical sizes of clusters seen in experiments.  To avoid
several of these complications, the vast majority of the theoretical work has been done on smaller clusters ( $<$ 20
atoms) where theoretical methods can be performed exhaustively.  For example, Malli and coworkers demonstrated the importance of relativistic effects on Au$_{13}$ clusters performing all electron calculations using the Dirac Scattered Wave method \cite{PRM89}.  Later work on Au$_{13}$ demonstrated that spin-averaged scalar relativistic theory was quite
adequate to reproduce the correct level spacings \cite{AC97} in all electron calculations.   Landman and coworkers have performed electronic structure calculations and geometry optimizations on gold cluster
anions of up to 14 atoms \cite{HYL3} using density functional theory (DFT), this time with a relativistic
norm-conserving pseudo-potential.  According to their calculations, clusters up to Au$_{12}^-$ have planar ground-state
geometries in agreement with experiments.  Further, they showed that all neutral clusters with even
numbers of atoms smaller than Au$_{14}$ have closed shells as evidenced by a gap between the lowest binding energy peak
and the rest of spectrum \cite{HL0}.  Having closed shells implies no magnetization.  This, as we will show later, does not hold in the case of larger gold clusters.  Wang and coworkers used the
LDA in the form of PW91 with a double zeta basis set and a scalar relativistic pseudo-potential to find lower
energy amorphous geometries relative to more symmetric ones \cite{WWZ2}.  They noted a variation in the HOMO-LUMO gaps depending on whether the number of atoms in the cluster was
even or odd.  The gap was bigger for an even number of atoms in the cluster.  This suggests a closed electronic shell and
that clusters with an even number of atoms are spin singlets.  None of these earlier calculations hinted at the
possibility that high spin ground states would also be possible. 

For larger clusters, a combination of semi-empirical and density functional theory has been popular.  Early work indicated that a truncated decahedral motif seems to include the most energetically favorable clusters \cite{CLS97,CLS97b}.  Subsequent geometry
optimizations of gold nano-particles of 38 and 55 atoms suggested that asymmetric non-bulk-like geometries are about 0.1
eV lower in energy than the more symmetric candidates \cite{GMB98,MRG99,FGB4}.  Hakkinen and coworkers showed  that Au$_{55}$ is icosahedral using hot
electron spectra of gold clusters.  All other anions near it are of lower symmetry types
\cite{HMK4}.  So far, most of the theory has been concerned with the shape and composition of the clusters and not on the details of the electronic structure and consequently the total spin.  A notable except is work by U. Landman et al. who investigated the detailed electronic structure of  bare and thiolated Au$_{38}$ using the local density approximation \cite{HBL99};  however, their work did not consider the spin-state of the gold cluster.  


In this paper, we use spin-dependent density functional theory in the scalar relativistic pseudo-potential
formalism to study the energetic tendency for the gold clusters to spin polarize. We will show explicitly how this
spin-polarization depends on the cluster size and that surface ligands are not needed to induce spin-polarization.  
We will show
through first principles calculations that gold clusters are intrinsically magnetic due to the hybridization of the atomic orbitals within
the scalar relativistic formalism.    This result is general for nano-sized clusters but does not apply to magnetism of gold surfaces which is likely induced by a different mechanism such as  chemiabsorption.   Additionally, we demonstrate that moderate changes in the cluster geometry do not overwhelm this magnetism, and it is plausible that this sort of magnetism will exist in more realistic amorphous geometries.

In section \ref{s:method}, we discuss the theoretical framework in which we model the electronic structure of these clusters.  In section \ref{s:results}, we present the main evidence that bare clusters tend to be spin polarized in a certain size regime, and in section  \ref{s:geometry}, geometry relaxation is seen to enhance the magnetization.  We conclude with a discussion of  implications of this result and foreshadow future work. 


\section{Magnetism from first principles}
\label{s:method}

The two main findings from the computational work thus far have been the need to account for relativistic effects and
the possibility of many energetically similar cluster geometries.  Relativistic effects are included in this work through
the use of a scalar relativistic pseudo-potential.  Perturbative addition of valence relativistic terms such as
spin-orbit, mass-velocity, and the Darwin term will be investigated in later work but are not necessary to reveal the underlying orbital-hybridization mechanics since the energy shifts, in general, are smaller than spin coupling.   

Using first principles to accurately determine the geometry is a more difficult problem \cite{CLS97,CLS97b}. While it is well known that DFT
tends to predict accurate geometries for simple molecules and bulk solids, little is known about the relative
trustworthiness of DFT for the structures of highly complicated molecules and large metal clusters.  Coupled with the
large errors possible from the choice of pseudo-potential, basis set, and functional, DFT in its current form may not provide a highly reliable geometries of gold clusters even at zero temperature.   However, DFT should provide reasonably accurate electron structure predictions.   Furthermore, gold clusters
must tend to an fcc bulk-like geometry in the large cluster limit since a large enough cluster will eventually look like
small piece of bulk gold.  This has been verified, for example, by Hori and coworkers from X-ray diffraction and Transmission Electron Microscopy  (TEM) for
gold nano-particles of 2.5 nm diameter.   We choose to use prototypes with high symmetry and to model our gold clusters as ideal cleaves from the bulk
face-centered cubic (fcc) gold lattice with a room-temperature  lattice spacing   
4.08 \AA.  This naive choice offers an additional advantage in that results for these cleaved gold clusters are
directly comparable to results in both the bulk and isolated atom limits. 

All results were generated using Gaussian 03 \cite{G03}
\footnote{Certain commercial software is identified in this paper in order to specify the  procedures adequately. Such identification is not intended to imply recommendation or endorsement by the National Institute of Standards and Technology, nor is it intended to imply that the materials or equipment identified are necessarily the best available for the purpose. }.  The basis sets were
chosen to be either single valence plus polarization with the Stuttgart effective core potential (SVP/STUTT) or the double zeta without polarization LANL2DZ with its
respective effective core potential.  Both have been used for transition metals with considerable success.  
The SVP  basis set consists of  27 basis functions comprised of 55 primitive gaussians.  The LANL2DZ basis consists of 24 basis functions comprised of 44 primitive gaussians.  In both cases, the pure 5d and 7f basis functions were used.

The density functionals we use are one generalized gradient approximation (PBE \cite{PBE96}) and two hybrid functionals
(B3LYP \cite{B93}, PBE1PBE ).  For the smaller clusters, the hybrid type functionals are perhaps the most likely to be
accurate since the clusters will be more molecule-like.  Hybrids are well trusted in chemistry and have become the standard for molecular
DFT-based calculations.   The hybrid functionals are known to converge more quickly because the longer range exchange
potential will widen the HOMO-LUMO gap.  PBE on the other hand is designed to be exact in the uniform gas limit,
and it should be expected to be reliable in the large cluster limit.  A combined analysis using both classes of functionals
is particularly necessary when considering larger molecules. 

For
Au$_{14}$ - Au$_{56}$ clusters, we used an ultrafine grid for density integrations, and a fine grid was used in the case of the Au$_{68}$ cluster.  
The ultrafine grid consists of 73 radial points and 43028 total points per atom, and the fine grid consists of 56 radial points and 17039 total points  per atom.
Throughout, we used tight
convergence criteria.  
Tight convergence means that the energy was converged to at least 1 mHartree and the RMS density matrix to less than 1 part in $10^{-8}$.
SCF convergence often required the use of extra quadratic convergence steps for the SVP/STUTT
basis.  
Geometry optimizations were constrained by symmetry, restricted to the spin-singlet state,  and performed using the B3LYP hybrid functional. 


\section{Spin-Polarized Bare Clusters}
\label{s:results}

For a single gold atom in the pseudo-potential formalism, 60 electrons are designated to be core electrons and the remaining
11 valence electrons form one S$_{1/2}$, four D$_{3/2}$ and six D$_{5/2}$ states.  

Therefore, we can expect that our treatment of relativistic effects through a pseudo-potential is adequate to reproduce most of the relativistic effects. 
D band splittings are reproduced by DFT  \cite{AC97}.


On the one hand, bulk gold is diamagnetic with a mostly filled d band  and s-shell  conduction electrons.  A small fraction of vacancies
in the d band is not sufficient to overcome the diamagnetism of the core electrons in the gold atoms \cite{clr4}.   In comparison in the atomic limit, the
d shell is completely full.  As the gold atoms aggregate into a cluster, the atomic shells will
hybridize.  As a result, the  d-shells develop vacancies when gold clusters increase in size.  The number of
vacancies will be proportional to cluster size.  At some intermediate size, the number of d shell
vacancies exceeds a critical threshold, and a spin polarized state will be the lowest energy one.  For larger
clusters, the core diamagnetism dominates, and this surface magnetism is suppressed.  How complete this suppression is
depends strongly on gold's magnetic screening length.

In order to test this theory of spin-polarization in the physical clusters, we consider highly idealized, isolated, bare, uncharged, single gold clusters with even
numbers of atoms.  (Clusters with an odd number of atoms will always result in paramagnetic clusters due to the unpaired
electron.)  Since experiments do not give the detailed cluster shapes, we build the theoretical clusters out of the known
face-centered cubic lattice of bulk Au. The clusters are constructed by picking atoms enclosed by spheres of a given
radius in the face-centered cubic bulk gold lattice.  This gives high-symmetry gold clusters with either an octahedral
shape (Au$_{14}$, Au$_{38}$, Au$_{68}$) or a bond-centered rod (Au$_{28}$, Au$_{56}$).  

For Au$_{38}$ and smaller clusters, we have been able to optimize the
geometry within the hybrid DFT method and find that uncapped clusters expand upon geometry relaxation.  The bare clusters are a good model for  alkylated clusters where there is a weak interaction between the ligand and the surface gold atoms.   The effects of strongly interacting ligands such as sulphur-based ones  will be
studied in later work.  The approximated geometries can be relaxed using the electronic structure results to optimize the
geometry; however, the choice of optimized or approximate geometries should not greatly affect the conclusions about
magnetism in these clusters.   Nevertheless, we test this proposition at the end of this section.

The approximate diameters for the clusters can be taken from measuring the distance from one surface gold atom to its
symmetry partner on the other side of the cluster.  For Au$_{14}$, Au$_{38}$,  and Au$_{68}$, this gives non-optimized diameters
of 0.71 nm, 0.92 nm, and 1.24 nm respectively.  Au$_{28}$ and Au$_{56}$ do not have spherical symmetry,  we report the largest surface to surface distance.  This gives non-optimized sizes
of 0.98 nm, and 1.19 nm respectively.  The reader is cautioned however that experimental nano-clusters
are capped and have charge distributions which can lead to a much larger effective radii.  Au$_{68}$ represents
the largest cluster for which we have done calculations in this work. 

\begin{figure}[ht]
\includegraphics[scale=0.2]{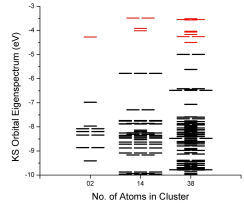}
\caption{Singlet orbital Kohn-Sham eigenspectrum (B3LYP/SVP/STUTT) for gold nano-clusters of various sizes.  The black lines are the occupied Kohn-Sham orbitals and the grey ones are the unoccupied ones.  Degenerate states are indicated by $n$ adjacent lines corresponding to the $n$ degrees of degeneracy.}
\label{f:eigs}
\end{figure}

For the gold dimer, we show the energy levels in the leftmost column of Figure \ref{f:eigs}.  The HOMO-LUMO gap 
is 2.1 eV.  The symmetry of the occupied orbitals is s-like.  The unoccupied orbital is
also s-like.  
The column on the right-hand-side of Figure \ref{f:eigs} depicts the corresponding energy levels for Au$_{38}$.   
The results in Figure \ref{f:eigs} indicate that  the unoccupied band gap vanishes in the case of the Au$_{38}$ cluster indicating that the spin singlet ground-state is unstable and prefers to decay to a spin polarized state. 
Figure \ref{f:eigs} demonstrates how the discrete eigenspectrum for gold evolves from the diatomic limit into a band-like spectrum for the
larger Au$_{38}$ cluster.  These eigenvalues are determined using the B3LYP functional to calculate the lowest spin ground-state (not necessarily lowest energy) on non-optimized geometries of neutral gold clusters.  

To find the spin state of a gold cluster, we compare the electronic energy of various spin constrained states. 
Neglecting possible spin-orbit coupling effects, the magnetic moment of a system is related to its spin by $M_z=2 \mu_B \langle S_z \rangle = 2 \mu_B \sqrt{S(S+1)}  $
where $\mu_B$ is the Bohr magneton and $S$ is the total spin quantum number.  The total spin moment, $M_z$, is this divided by
the number of atoms, $n$, in a cluster ($m= M_z/n$), and then $m$ is commonly reported in the experimental literature.  
The experiments are performed
at finite temperature and our calculations are at zero temperature.  Table \ref{t:moments} shows the magnetic moment per atom versus spin state for several clusters we studied.   On the left, we list several  possible numbers of unpaired electrons for the neutral bare clusters.   We arbitrarily stop at 10 unpaired electrons.  This choice as we shall present later is motivated by our theoretical results that show that it is energetically unlikely for the gold nano-particles to become so strongly polarized.  Given these values for the numbers of unpaired electrons, the table presents a range of physically possible magnetic moments.  The experimentally suggested value of 0.036 $\mu_B/$atom  is slightly smaller than the weakest polarized values for each cluster type listed here.   

\begin{figure}[ht]
\includegraphics[scale=0.2]{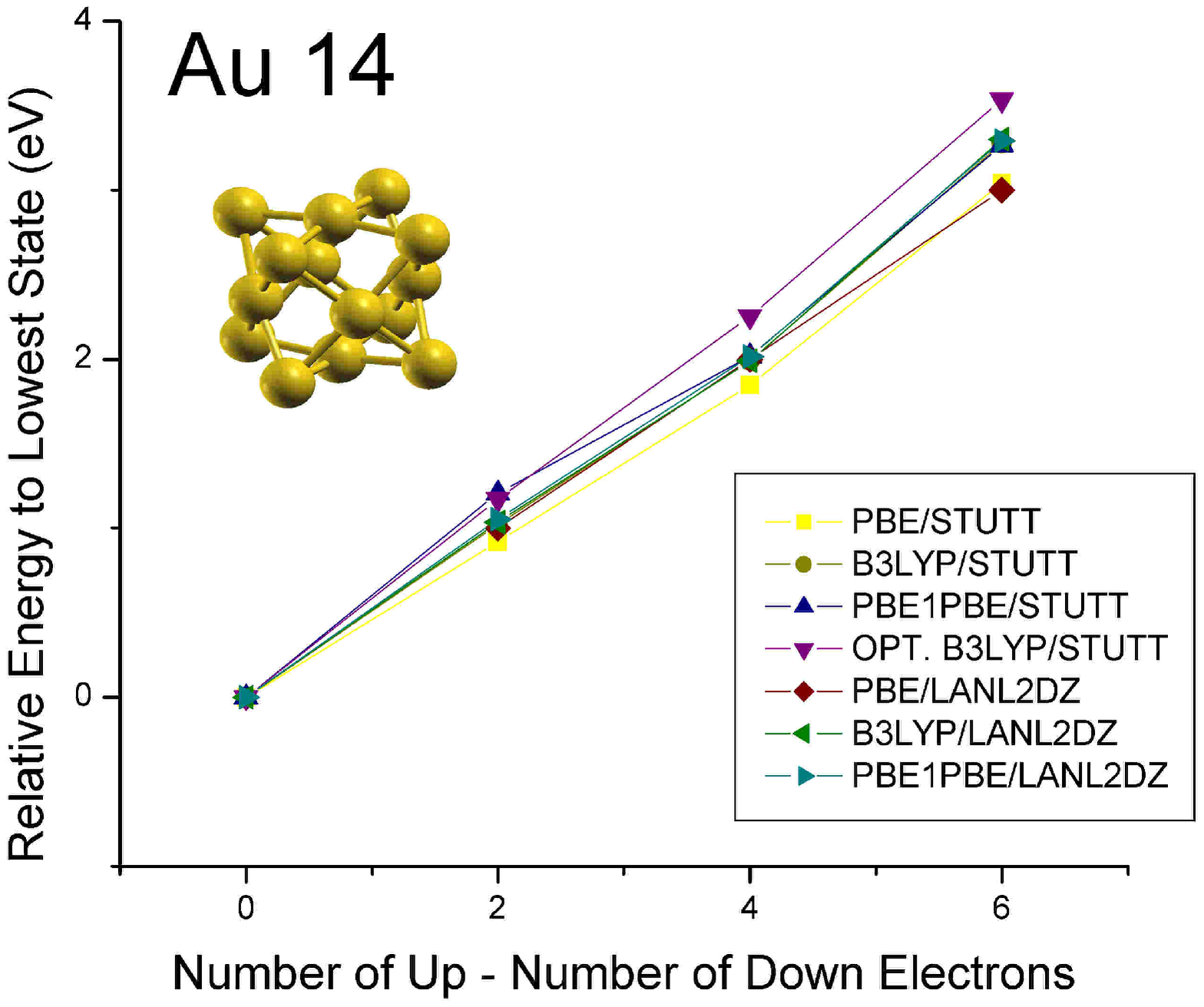}
\caption{Total energy for the gold Au$_{14}$ nano-cluster as a function of the spin state as described via various density functional methods.}
\label{f:etot14}
\end{figure}

Figure \ref{f:etot14} shows a plot of the energy for Au$_{14}$ relative to the lowest-energy electronic spin-state.  We find the energetic minimum is spin unpolarized (singlet).   This seems to indicate that small bare gold clusters are, as expected, diamagnetic.   It also agrees with
the naive picture of complete d orbital filling and unbroken symmetry between different spin species.  Geometry
optimization expands this cluster but does not change the qualitative spin state.  This cluster has a very large surface
to volume ratio with all atoms being surface atoms.  Furthermore, all the surface atoms have the same coordination
number.  It is energetically unfavorable to have holes or localized moments in the d band.  A similar results holds for
Au$_{28}$ (Not shown). 

\begin{figure}[ht]
\includegraphics[scale=0.2]{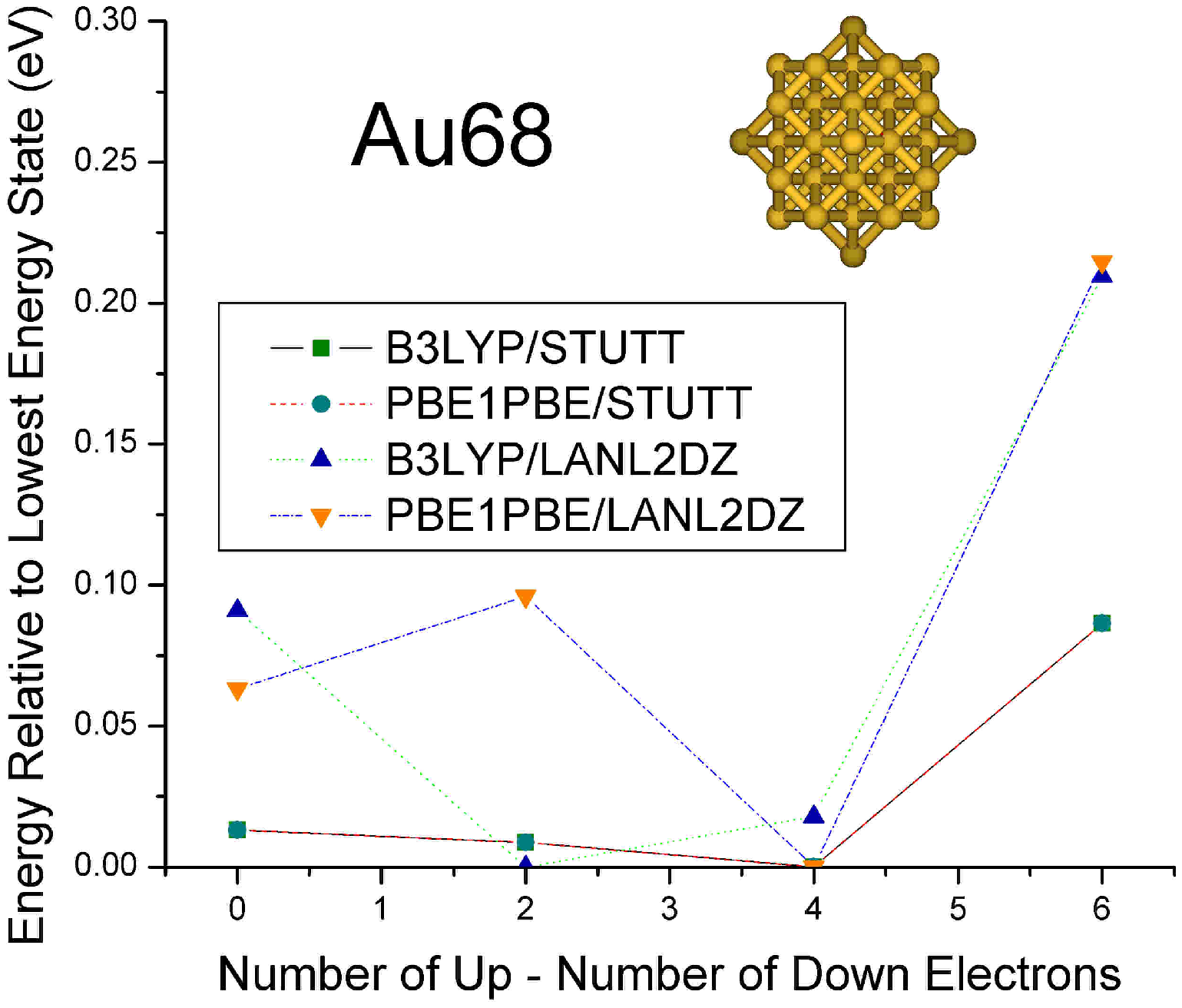}
\caption{Total energy for the gold Au$_{68}$ nano-cluster as a function of the spin state as described via various density functional methods.}
\label{f:etot68}
\end{figure}

For intermediate sized clusters, a different story unfolds.  Figure \ref{f:etot38} depicts the relative  energy for Au$_{38}$  for various spin polarizations.  We find the energetic minimum ($-0.2$ eV relative to the spin singlet) is spin triplet.   This minimum is significant even at room temperature of 300 K where $k_B T = 0.03$ eV.  Previous calculations on this cluster showed a six-fold degenerate HOMO.  Here, we use hybrid functionals, a slightly higher level of theory, and find that the ordering of orbitals has shifted for this cluster, and the HOMO is no longer six-fold degenerate.   The spin polarized state arises because the atomic orbitals hybridize, and it becomes
energetically favorable for holes to form.    Geometry optimization enhances the energetic tendency to
hybridize and consequently to polarize.  The trend is most pronounced for the PBE functional but also evident in the hybrid functionals.  For the larger octahedral cluster, Au$_{68}$, we obtain a similar total energy curve (Fig. \ref{f:etot68}).  In this case, the energy minimum is only $-0.1$ eV lower than the singlet, but the lowest energy state is likely a spin quintuplet.  
Given the prohibitive computational expense involved in the use of the ultrafine integration grids for our DFT calculations, a smaller grid was used in the case of Au$_{68}$.  



\begin{figure}[ht]
\includegraphics[scale=0.2]{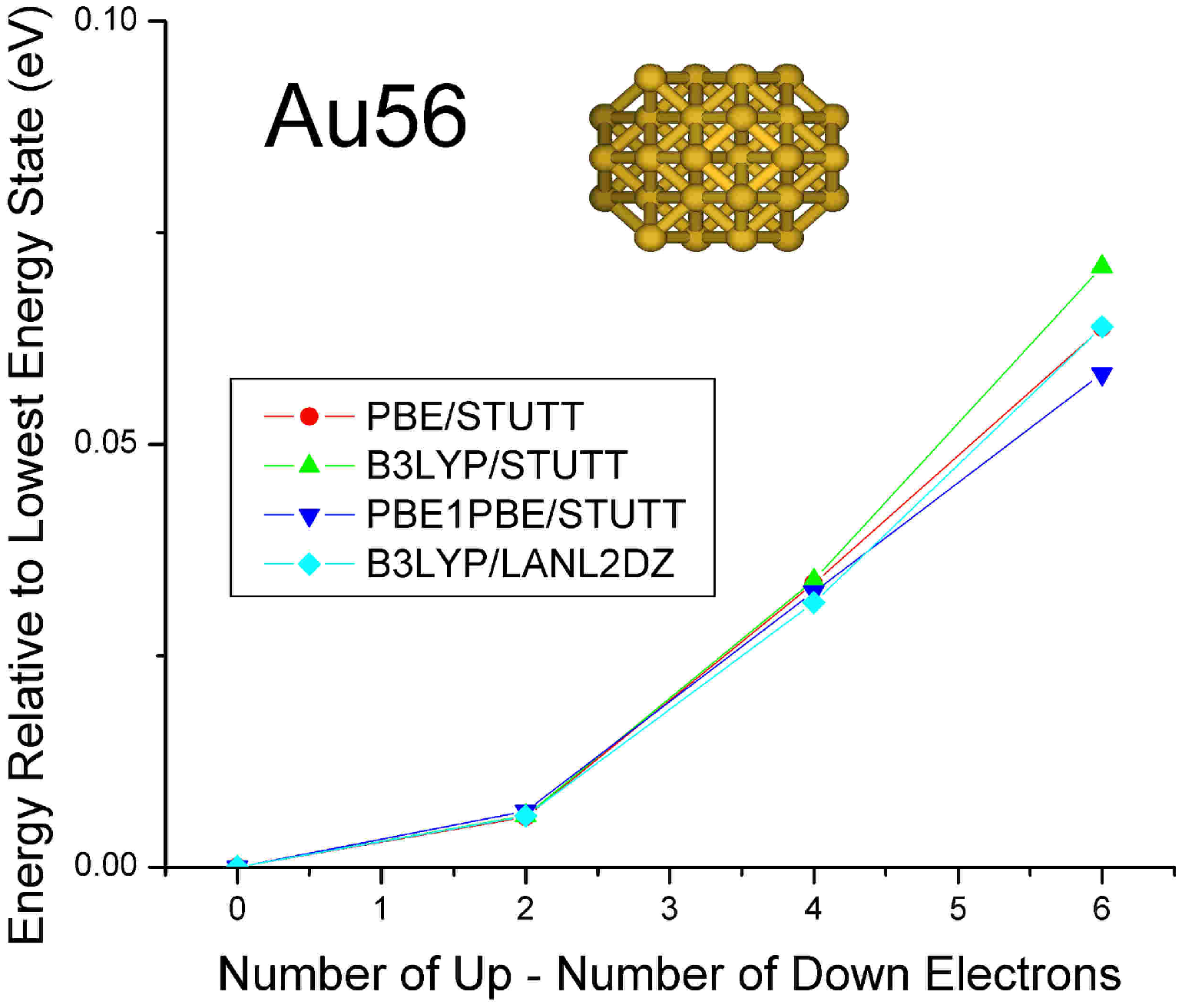}
\caption{Total energy for the gold Au$_{56}$ nano-cluster as a function of the spin state as described via various density functional methods.}
\label{f:etot56}
\end{figure}

Figure \ref{f:etot56} shows the corresponding  energy plot  for Au$_{56}$.  This is a bond-centered cluster and should not be expected to be directly comparable to the same  sized octahedral shaped cluster.  However, the trends within the set of bond-centered clusters should be similar to the trends seen in the octahedral clusters.  The spin singlet and triplet states are degenerate within our level of theory.  For Au$_{28}$, we obtained a spin singlet (not shown), so while this cluster is not large enough to be polarized, it is closed to the threshold size at which spin polarization arises. 

\begin{figure}[ht]
\includegraphics[scale=0.1]{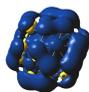}
\caption{The spin polarization for an Au$_{38}$ gold nano-cluster (B3LYP/LANL2DZ).    The majority-spin polarization is mostly localized to the surface of the cluster.  The diamagnetic core is obstructed by the spin-polarized surface.}
\label{f:spindens38}
\end{figure}

The electronic spin polarization for the lowest energy (triplet) state of Au$_{38}$ (B3LYP/LANL2DZ) has been plotted in Figure \ref{f:spindens38}.  The spin polarization is mostly localized on a
surface mono-layer of the cluster.  
A Mulliken analysis of the spin density shows an average spin of -0.0706 on each of the 6 interior atoms and an average of 0.0757 on each of the 32 surface atoms.  The distribution of spin moments is not uniform on the surface, but rather the greatest spin-polarization occurs along a plane perpendicular to the z-axis.  The strong localization to the surface suggests that the magnetism is a
surface effect and is likely to be quenched by a strong interaction with capping ligands such as sulfur.  A overall interior polarization opposes the strong surface effect.  This is consistent with the idea that bulk gold is diamagnetic and that the diamagnetism per atom is weaker than the tendency of the surface atoms to polarize.  The
core tends to oppose, although weakly, the magnetization of the surface which is clearly a size-dependent phenomenon.  

A similar result is found for the Au$_{68}$ (quintuplet).  The spin density plot is not shown.  The surface has 8 groups of three gold atoms and 6 single atom outliers for a total of 30,  all with majority spin.   The outliers have the highest local spins.
The semi-interior  atoms (16 atoms) are partially screened from the exterior  and have small  spins.  The interior consists of the 14 atom octahedron and the 8 atoms which cap their faces.   The core is diamagnetic with the strongest moments on the 8 capping atoms.  The Mulliken populations of the spin density show an average spin of -0.103 on each of the 14 interior, 0.002 of each of the semi-interior atoms, and 0.180 on each of the 30 surface atoms.

For both of the spin-polarized clusters, these results suggest  a paradigm of a spin-polarized paramagnetic surface and an anti-polarized diamagnetic core.    




\section{Geometry-Induced Spin Polarization Enhancement}
\label{s:geometry}

For Au$_{38}$ and Au$_{14}$, we have optimized the cluster geometry within the B3LYP/LANL2DZ formalism.    Geometry optimization for clusters of these sizes tend to be very expensive.  For this reason we optimized only the singlet ground state and invoked symmetry.  For non-singlets the geometry will vary slightly; however, it is unlikely that the difference among the various spin-state geometries will be greater than the the difference between the initial non-optimized geometry and the optimized singlet one.  Our initial coordinates are the octahedral coordinates from the face-centered cubic (fcc) gold lattice at room temperature and are expected to be slightly expanded relative to coordinates as zero temperature.  First principle 
calculations are implicitly for an isolated cluster at zero temperature.    

We use two measures to compare the relative geometry changes for our clusters.  First is the average RMS displacement of a gold atom from the center of mass of the cluster.  This can be calculated for both the non-optimized and optimized clusters.   The second measure is the furthest surface to surface distance.  This gives a rough measure of the diameter since the physical diameter depends on the electronic charge distribution and is not well defined.    

\begin{figure}[ht]
\includegraphics[scale=0.2]{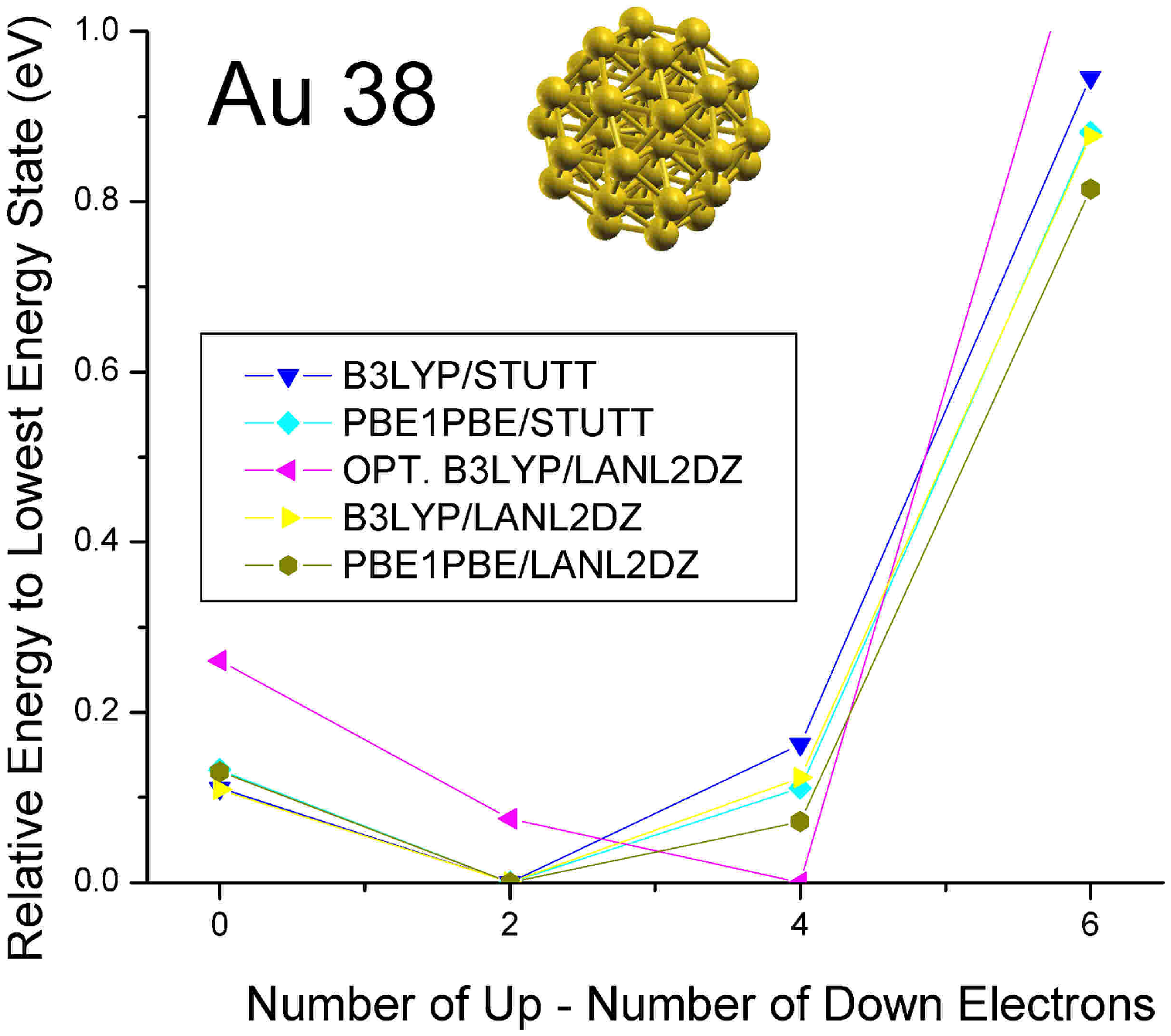}
\caption{Total energy for the gold Au$_{38}$ nano-cluster as a function of the spin state as described via various density functional methods.}
\label{f:etot38}
\end{figure}

Au$_{38}$ was the only spin-polarized cluster for which we were able to perform geometry optimization.  Electronic
structure calculations performed on the optimized geometries shows curves (Fi. \ref{f:etot38}) with slightly deeper wells.   
Since we did not constrain the geometry during optimization the clusters shifted slightly from the expected octahedral geometries.  In order to quantify the relative sizes of the clusters prior and after geometry relaxation, we compare the average root-mean-squared distance from the center of the cluster.  

In our calculations, both clusters expand upon geometry relaxation.    This is in contradiction to what is often cited in the literature \cite{ZTA0,CLS97,CLS97b}, in which a 1\% reduction in size relative to the bulk was observed (EXAFS).  However, in the reference, the experiment was conducted on thiol-capped nano-particles.    Later measurements by Crespo et al. on Thiol capped nano-particles suggest that they in fact do expand from 2.83 to 2.98 \AA \cite{HBL99}.  

For Au$_{14}$ the average RMS value goes from 2.701 to 2.917 \AA.  This is an 8\% increase in size and is consistent with the 5\% experimental increase observed by Crespo et al.   Moreover, there is considerable geometry distortion for the small gold cluster so while the average RMS displacement expands, the longest surface to surface distance decreases from 7.067  to 6.888 \AA.   The optimization and distortion  have little if any effect on the magnetic properties of the 14 atom cluster as seen in Fig. \ref{f:etot14}.   

For Au$_{38}$, we see further evidence of cluster expansion.  The RMS displacement expands from 3.947 to 4.066 \AA ~ which gives a net expansion of  3\%.  This result also is in reasonable agreement with Crespo's measurements.  Considering that strongly-interacting ligands such as thiolated ones are expected to cause expansion relative to the bare cluster, the ligands are thought to further expand the clusters and this too would agree with the experimental results.  
This cluster has six (100) facets and eight (111) facets.  As also seen in previous LDA optimizations of this cluster \cite{HBL99}, the atoms in the center of the (111) facets move outward to make the cluster more spherical.  However, the final bond lengths after optimization with the hybrid functional are 5\% longer and in the range  2.83-2.95 \AA.  This is to be expected since LDA typically over-binds systems and functionals such as Hybrids that account for long-ranged exchange correct this.  

Despite geometry changes, the magnetic spin state is predicted to be the ground d state at both contracted and expanded geometries.  In fact, our data indicate that expansion favors a more strongly magnetized ground-state perhaps by encouraging greater hybridization.  The furthest distance among surface atoms expands as well in this case from 9.123 to 9.235 \AA.     The Au$_{38}$ cluster exhibits less distortion than the smaller one due to the inner 14 gold atoms retaining the octahedral structure.   
Overall, as observed in Fig. \ref{f:etot38}, full geometry optimization leads to larger magnetization in the cluster indicating that the conclusions regarding the rise of magnetic behavior in these clusters are not significantly affected by geometry optimization.




\section{Conclusions}

Density functional calculations presented in this work show that magnetism is, in fact, intrinsic to gold clusters of a nano-sized regime due to atomic orbital hybridization.  The magnetism is localized to a mono-layer on the outside of these
clusters with the interior atoms remaining mostly diamagnetic.   These calculations support a paradigm of a spin-polarized paramagnetic surface and an anti-polarized diamagnetic core.  This arrangement allows for a leveling off of the magnetic
moment per atom as the clusters increase in size.  For larger clusters the diamagnetic core will dominate as the surface
to volume ratio becomes small.   Thus, magnetism in gold clusters is a primarily a size dependent effect.  The calculations are consistent with the experimental evidence that suggests that gold magnetism is strongest
when weakly interacting capping agents are used.    Strongly interacting capping agents would likely quench this magnetic behavior.   
Future work will examine how sulfur and nitrogen based surface ligands affect this result.  Preliminary calculations
indicate that sulphur does quench the spin polarization by reinforcing the energetic stability of the singlet state.
Geometry optimizations within GGA and Hybrid density functional theory indicate that the average inter-atomic spacings in gold clusters tend to expand relative to the bulk value.   However, these geometry changes do not change the qualitative result that bare gold nano-clusters are magnetic in certain size regimes.   
The next step in our investigation will involve the use of semi-empirical tight binding formulations to obtain  better structures followed by single point electronic structure calculations.  Since these systems are considered at finite temperature dynamics must also be considered.







\begin{table}
\begin{center}
\caption{The magnetic moment per atom for different spin states in gold nano-clusters.}
\label{t:moments}
\begin{tabular}{l|c|c|c|c|c|c}\hline
Number of Up - Down Electrons       & Au$_{2}$ &  Au$_{14}$ & Au$_{28}$ & Au$_{38}$ & Au$_{56}$   & Au$_{68}$ \\
2     &  1.414  & 0.202 & 0.101 & 0.074 & 0.051 & 0.042 \\
4     &  2.449  & 0.350 & 0.175 & 0.129 & 0.087 & 0.072 \\
6     &  3.464  & 0.495 & 0.247 & 0.182 & 0.124 & 0.102 \\
8     &  4.472  & 0.639 & 0.319 & 0.235 & 0.160 & 0.132 \\
10   &  5.477  & 0.782 & 0.391 & 0.288 & 0.196 & 0.161 \\
\end{tabular}
\end{center}
\end{table}

\end{document}